\documentclass{article}
\usepackage{ijcai15}
\usepackage{times}
\usepackage{amsmath,amssymb}
\usepackage{mathrsfs}
\usepackage{amsthm}
\newtheorem{example}{Example}

\usepackage{tikz}
\newcommand{\cnum}[1]{\tikz\node[draw,circle,inner sep=1pt] {{\scriptsize\rm #1}};}
\newcommand{\Qesto}{\textsc{Qesto}\xspace}
\usepackage{xspace}
\usepackage[linesnumbered,ruled,vlined]{algorithm2e} %
\SetKwFunction{SAT}{SAT}
\SetKwFunction{cnf}{CNF}
\SetKwFunction{id}{ID}
\SetKwFunction{analyze}{Analyze}
\SetKwFunction{max}{max}
\SetKwFunction{var}{var}
\SetKwFunction{InitCondition}{InitCondition}
\SetKwData{round}{qlev}%
\SetKwData{conflict}{conflict}%
\SetKwData{btround}{btlev}%
\SetKwData{out}{outc}%
\SetKwData{nul}{$\bot$}%
\SetKwData{false}{false}
\SetKwData{true}{true}
\SetKw{Goto}{Goto}
\SetKwInOut{Input}{input}
\SetKwInOut{Output}{output}
\DontPrintSemicolon%
\SetKw{Func}{Function}

\def\miko{Mikol\'a\v{s} Janota}
\def\jpms{Joao Marques-Silva}
\def\theTitle{Solving QBF by Clause Selection}
\title{\theTitle\thanks{
This work is partially supported by
SFI~PI~grant~BEACON (09/IN.1/I2618),
FCT grant POLARIS (PTDC/EIA-CCO/123051/2010),
CMU-Portugal grant AMOS CMUP-EPB/TIC/0049/2013,
and national funds through Funda\c{c}\~ao para a Ci\^encia e a Tecnologia (FCT) with
reference UID/CEC/50021/2013.
}}

\author{\miko$^{1}$\and\jpms$^{1,2}$%
\\$^1$\,INESC-ID, IST, Lisbon, Portugal
\\$^2$\,University College Dublin, Ireland
\\\url{mikolas@sat.inesc-id.pt},
  \url{jpms@tecnico.ulisboa.pt}
}
\newcommand{\restrict}[2]{{#1}|_{#2}}
\newcommand{\comprehension}[2]{\ensuremath{\left\{ {#1} \;|\; {#2}\right\}}}
\newcommand{\restr}[2]{{#1}|_{#2}}
\DeclareMathOperator*{\lev}{{\sf lv}}
\DeclareMathOperator*{\condition}{\mathscr{C}}
\usepackage{url}
\begin{document}
\maketitle
\begin{abstract}
  Algorithms based on the enumeration of implicit hitting sets find a
  growing number of applications, which include maximum satisfiability
  and model based diagnosis, among others.
  This paper exploits enumeration of implicit hitting sets in the
  context of Quantified Boolean Formulas (QBF). The paper starts by
  developing a simple algorithm for QBF with two levels of
  quantification, which is shown to relate with existing work on
  enumeration of implicit hitting sets, but also with recent work on
  QBF based on abstraction refinement.
  The paper then extends these ideas and develops a novel QBF
  algorithm, which generalizes the concept of enumeration of implicit
  hitting sets.
  Experimental results, obtained on representative problem instances,
  show that the novel algorithm is competitive with, and often
  outperforms, the state of the art in QBF solving.
\end{abstract}
\section{Introduction}

The enormous success of SAT enabled us to solve problems that go beyond the NP
complexity class.  Among these, Quantified Boolean Formulas (QBFs) represent
an important formalism. Indeed, as deciding QBFs is PSPACE-complete, they
cover a wide range of problems.  Namely from model checking, planning, or
two-player games~\cite{Papadimitriou-94,Benedetti-jsat08,Rintanen07,Narodytska-cav14}.

In the recent years, a number of approaches were developed to solve QBF.
Among the most representative are \emph{conflict/solution driven
learning}~\cite{cadoli-aaai98,rintanen-ijcai99,Zhang-iccad02,giunchiglia-ijcai01},
which extends SAT clause learning, and a number of approaches that \emph{expand} the given
QBF into a SAT problem~\cite{Biere04,Benedetti04,janota-sat12}

This paper proposes a novel approach for QBF solving that has two distinctive
features: 1)~it uses a SAT solver as an NP oracle,
2)~it is anchored in the duality hitting set principle.
In the Boolean CNF world, the duality hitting set principle  lets us
reason about \emph{minimally unsatisfiable sets (MUSes)}
by traversing \emph{maximally satisfiable sets (MSSes)}.
This has numerous applications in MUS computation (cf.\,\cite{reiter-ai87,liffiton-jar08}) and more recently in MaxSAT
solving and other domains~\cite{karp-cpm10,karp-soda11,bacchus-cp11,stern-aaai12,karp-or13,bacchus-sat13,pms-aaai13,bacchus-cp13,slaney-ecai14}.

We first relate MUS computation to solving QBF with two levels of
quantification ($\forall\exists$). Then, we extend this idea for arbitrary
number of quantification levels. Hence, the proposed approach is not only a
novel approach to solving QBF but it can also be seen as an extension of the
duality hitting set  principle for PSPACE.

The proposed approach instantiates a SAT solver at each quantification level,
whose task is to select or deselect clauses at that level. The fact that we
are using a SAT solver as the underlying reasoning engine in our
algorithm is a big engineering advantage. Indeed, if a better SAT solver
appears, it can simply substitute the old one in our implementation.
The main contributions of the paper are summarized as follows:

  (1)~A novel algorithm for solving QBF, called \Qesto, is developed.
  (2)~A link is established to implicit hitting set algorithms.
  (3)~A link is established to the CEGAR-based QBF solving.
  (4)~A prototype was implemented and evaluated.

  The paper begins by overviewing the concepts and notation used (Sec.~\ref{sec:preliminaries})
  and continues by developing an algorithm for two-level QBF (Sec.~\ref{sec:2qbf}).
  Sec.~\ref{sec:qbf} generalizes this approach to general QBF
  and Sec.~\ref{sec:others} relates to other concepts and algorithms.
  Finally Sec.~\ref{section:results} describes experimental evaluation
  and Sec.~\ref{sec:conclusions} concludes the paper.

\section{Preliminaries}\label{sec:preliminaries}

Throughout the paper we assume an infinite set of Boolean variables, commonly denoted as $x$, $u$,
$e$, etc. Boolean connectives ($\land$,$\lor$,$\lnot$) are used under the traditional semantics.
An assignment to a set of Boolean variables $X$ is a mapping from $X$ to the Boolean constants  $\{0, 1\}$.
For an assignment $\tau\,:\,X\rightarrow\{0,1\}$
and a formula $\phi$ we write $\restr{\phi}{\tau}$ for  the \emph{application} of $\tau$
to the formula $\phi$. Hence $\restr{\phi}{\tau}$ is obtained by
replacing all occurrences  of each variable $x\in X$ by the constant~$\tau(x)$
and any trivial simplifications are performed ($y\lor 1\equiv 1$, etc.).
An assignment~$\tau$ is a \emph{satisfying assignment} for a formula $\phi$ if
$\restr{\phi}{\tau}=1$. A formula is called \emph{satisfiable} if there exists a satisfying assignment
for it and  otherwise it is \emph{unsatisfiable}.
A satisfying assignment of a formula $\phi $ is often referred to as a \emph{model} of $\phi$.
For two formulas $\phi $ and $\psi $ we write $\phi\models\psi $ if any satisfying
assignments of $\phi $ is also a satisfying assignment of $\psi $.

A Boolean formula is called a \emph{literal} if it is a Boolean variable or its negation.  
For a literal $l$, we write $\var(l)$ for the variable in $l$.
A Boolean formula is called a \emph{clause} if it is a disjunction of literals.  A Boolean formula is
in Conjunctive Normal  Form (CNF) if it is a conjunction of clauses.  As common in literature,
we often treat clauses as sets of literals (implicitly disjoined) and formulas in CNF as sets of
clauses (implicitly conjoined). Observe that the empty set of clauses is semantically true (as the empty
conjunction is true) and the empty clause is semantically equivalent to false (as  the empty
disjunction is false).

Quantified Boolean formulas (QBFs)~\cite{GiunchigliaMN09}  extend Boolean formulas by allowing to quantify over variables
using the existential or the universal quantifier ($\exists$,~$\forall$).  Any QBF can be rewritten
to an equivalent Boolean formula by rewriting any
$\forall x.\,\Phi$ as
$\restr{\Phi}{\{x\mapsto 0\}}\land\restr{\Phi}{\{x\mapsto 1\}}$
and
$\exists x.\,\Phi$ as
$\restr{\Phi}{\{x\mapsto 0\}}\lor\restr{\Phi}{\{x\mapsto 1\}}$.

A QBF is \emph{closed} if all variables are bound by a quantifier.  A QBF is in \emph{prenex form}
if it can be written as $Q_1 x_1\dots Q_n x_k.\,\phi$ where $\phi $ is a Boolean formula and
$Q_i\in\{\exists,\forall\}$.  For formulas in prenex form, the quantifier part is called the
\emph{prefix} and the Boolean part the \emph{matrix}.  Unless specified otherwise, in this
paper we will be dealing with formulas that are both prenex
and closed and the matrix is in  CNF. Any adjacent variables with the same quantifier in the
prefix are merged into \emph{blocks}. This enables us to write QBFs in the form $Q_1 X_1\dots Q_n
X_k.\,\phi$ where $X_i$ are sets of variables and $Q_i\neq Q_{i+1}$ is assumed. We say that $X_i$
is a block at the \emph{quantification level}~$i$. We write $\lev(x)$ for the \emph{quantification
level} of a variable $x$, i.e., $\lev(x)=i$ if $x\in X_i$. We extend this notation to literals.
If a variable $\lev(x)=i$ and $Q_i=\forall$ than we say that~$x$ is universal and
analogously for existential variables.
The class of closed QBF formulas in prenex form with CNF matrix is referred to as
PQCNF.

We model a SAT solver by the function $\SAT(\psi)$, which returns a
satisfying assignment if the given formula~$\psi$ is satisfiable; otherwise it returns the
constant $\bot$.

An important fact about PQCNF formulas is that any
variables universally quantified at the end of the prefix can be removed.
For instance, $\exists e_1e_2\forall u.\,(e_1\lor u)\land(\lnot e_2\lor\lnot u)$
is equivalent to
$\exists e_1e_2.\,e_1\land\lnot e_2$.
\section{Solving Two-level PQCNF}\label{sec:2qbf}

Consider a formula $\forall X\exists Y.\,\phi$, where $\phi$ is in CNF. To
motivate our discussion, let us investigate what must happen for the formula
to be \emph{false}. There must be an assignment~$\tau$ to the variables $X$ so that
there is \emph{no} corresponding value for the~$Y$ variables that would make the
matrix~$\phi$ true. In other words, the CNF~$\restrict{\phi}{\tau}$ is
\emph{unsatisfiable}.  In the opposite case, when the formula is \emph{true}, any
assignment~$\tau$ to~$X$ makes the CNF~$\restrict{\phi}{\tau}$ satisfiable. Hence,
to decide whether a given 2-level PQCNF formula is true or not, it is sufficient
to decide whether there exists such assignment to~$X$ or not. This is the approach we
will take from now on.

Now let us look more closely at the effect of the application of~$\tau$ to~$\phi$.
Consider the simple example $\forall u\exists e.\,\{(u\lor e),(u\lor \lnot
e),(\lnot u\lor e)\}$. If $\tau(u)=1$, the application of $\tau$~yields the CNF
$\{(e)\}$ and $\tau(u)=0$ yields the CNF $\{(e),(\lnot e)\}$.  We can see that the
application of~$\tau$ acts as a \emph{selector} on the clauses comprising of
existential variables. In this particular case, $\tau(u)=0$ selects an
unsatisfiable set of clauses.  This resembles the way clauses are
selected in recent approaches for implicitly manipulating hitting
sets~\cite{karp-soda11,bacchus-cp11,stern-aaai12,karp-or13,pms-aaai13}.
However, in QBF, clauses cannot be selected arbitrarily. Consider
another simple example
$\forall u\exists e.\{(u\lor e),(\lnot u\lor \lnot e)\}$.
Here, regardless of the value of~$u$, the clauses are never selected at
the same time, i.e., the result is always satisfiable. This is because selecting a
clause requires setting to false all of its universal literals.

The above ideas give rise to a natural algorithm.  First, select some set
of clauses $S$ and test for its satisfiability. If it is
unsatisfiable, stop. If it is satisfiable, block~$S$ and any of its
subsets to be selected in the future and then repeat. The blocking step can be improved by looking
at the satisfying assignment~$\mu$ of $S$. Such assignment  satisfies some set~$S'$
s.t.\ $S\subseteq S'$.  Hence, we can block any
set that is a subset of~$S'$.

\begin{algorithm}[t]
\caption{Two-level Selection QCNF Solving}\label{algorithm:2qbf}
\Input{$\forall X\exists Y.\,\phi$}
{\bf output}:{truth value}
$\condition\gets\comprehension{(\lnot s_C\lor\lnot l)}{C\in\phi,l\in C, l\text{ universal}}$\;
\While{\true} {
$\tau\gets\SAT(\condition)$%
\label{step:CNFcandidate}%
\tcp*[r]{compute selection}
\lIf{$\tau=\nul$}{\Return \true}
$S\gets\comprehension{C\in\phi}{\tau(s_c)=1}$\;
$\phi'\gets\comprehension{\bigvee_{l\in C,\var(l)\in Y}l}{C\in S}$\;
$\mu\gets\SAT\left(\phi'\right)$\label{step:CNFcounterexample}\tcp*[r]{find a model}
\lIf{$\mu =\nul$}{\Return \false\label{step:CNFsolution}}
$S'\gets\comprehension{C\in\phi}{\text{ there is } l\in C \text{ s.t.\ }\mu(l)=1}$\;
$\condition\gets\condition\land\bigvee_{C\in\phi,C\notin S'} s_C$\;
}
\end{algorithm}

This idea is made precise by Algorithm~\ref{algorithm:2qbf}.
We consider a  formula $\condition$, which captures the requirements on
the considered  sets of clauses.  Sets of clauses are modeled by \emph{selection variables}
$s_C$, which have the traditional meaning that a clause $C\in\phi$ is
selected whenever $s_C$ is true. The formula $\condition$ is initialized by the
condition that a clause can be selected only if it is possible to set all of its
universal literals to false.

The algorithm iterates over the following sequence
of steps. First it chooses some set of clauses that is permitted by~$\condition$
(ln.~\ref{step:CNFcandidate}). If no such set exists, the algorithm terminates
and returns true (the formula is true because there is no set of clauses  yielding
unsatisfiability). Next,  the algorithm tests whether the existential part of the selected
clauses are satisfiable (ln.~\ref{step:CNFcounterexample}). If it is
unsatisfiable, then we are done because~$\tau$ is an assignment to the universal
variables that selects an unsatisfiable set of clauses.
If not, we block all subsets of clauses satisfied by the model.
This is done by calculating the set $S'$ of clauses satisfied by the assignment to the existential
variables and then requiring that at least one clause from its complement is selected.

\begin{example}
  Consider the following matrix with the prefix $\forall u_1u_2u_3\exists e_1e_2e_3$;
  the clauses are identified by a number on the left and they are graphically split into the
  universal and existential part:
\[
\begin{array}{cccc}
  \cnum{1}& u_1\lor u_2&\lor & e_1 \lor e_2 \\
  \cnum{2}& u_1\lor u_3&\lor & \lnot e_1 \lor e_2 \\
  \cnum{3}& \lnot u_1\lor u_2&\lor & e_3\\
  \cnum{4}& \lnot u_2\lor u_3&\lor & \lnot e_1\lor e_3\\
\end{array}
\]

Let us begin by selecting the clauses \cnum{1} and \cnum{2},
i.e., \mbox{$s_1=s_2=1$}. This is possible because it is
possible to set to false all the literals $u_1$,~$u_2$,~$u_3$.
The selected clauses correspond to the existential part $\phi'=\{e_1\lor e_2,\lnot e_1\lor e_2\}$,
which
is satisfiable by the assignment $\mu=\{e_1\mapsto 0,e_2\mapsto 1,e_3\mapsto 0\}$, let's say.
The assignment~$\mu$ satisfies the clauses \cnum{1}, \cnum{2}, and \cnum{4}\hspace{1pt}. Hence,
the clause \cnum{3} must be selected in
any further attempts, i.e., the condition~$\condition$ is strengthened by the unit clause $(s_3)$.
Due to this condition,
in the next iteration the algorithm \emph{cannot} select the clause \cnum{1} nor \cnum{2}
since they contain the literal~$u_1$ but \cnum{3} contains the literal $\lnot u_1$.
Let's say the algorithm selects \cnum{3} and \cnum{4}, which yields the existential part
$\phi'=\{e_3,\lnot e_1\lor e_3\}$, satisfiable by
$\mu=\{e_1\mapsto 0,e_2\mapsto 0,e_3\mapsto 1\}$,
which satisfies the clauses \cnum{2}, \cnum{3}, and \cnum{4}.
This means that the clause \cnum{2} must be selected in any next attempts,
which is a contradiction with the previous condition because \cnum{2} and \cnum{3}
cannot be selected at the same time due to the respective literals $u_1$ and~$\lnot u_1$.
Hence the algorithm terminates and responds that the formula is true.
\end{example}
\section{Generalizing for Arbitrary Prefixes: \Qesto}\label{sec:qbf}

Now we look at how the ideas for solving PQCNF by selection
generalize to prefixes of arbitrary length. Hence, we assume that  we
are given a formula of the form $Q_1X_1\dots Q_nX_n.\,\phi$, where
$Q_i\in\{\exists,\forall\}$, $Q_n =\exists$, and~$\phi$ is a propositional formula in CNF.
Recall that~$Q_n=\exists$ can be assumed
as any universal variable at the end of the prefix can be removed.

For the sake of presentation, we adopt the \emph{game-theoretic} point of view
on QBF (cf.~\cite{Goultiaeva-ijcai11}). Any QBF in prenex closed form is seen as a game between the
\emph{universal} and \emph{existential player}. The universal player tries to
falsify the matrix while the existential tries to satisfy it. Under this
perspective, a QBF is true if and only if there is a \emph{winning strategy} for the
existential player.

More precisely,  a game consists of~$n$ rounds and  in round~$i$, the player~$ Q_i$ assigns a value
to all the variables~$X_i$. At the end of the game we evaluate the matrix $\phi$ by the
accumulated assignment. The game is won by the universal player if the matrix evaluates to false,
and it is won by the existential player  if it evaluates to true.  We say that there is a winning strategy
for the existential player if the player can play so that he wins any game irregardless of how the
universal player  plays. Analogously, we define a winning strategy for the universal player.  A
formula is true if and only if there is a winning strategy for the existential player.
Conversely, formula is false if and only if there is a   winning strategy for the universal player.
\begin{example}
  Consider the formula $\forall u\exists e.\,\{\lnot u\lor e,u\lor\lnot e\}$.
  The assignments $\{u\mapsto 0\}$,$\{e\mapsto 1\}$ represent a game where the
  existential player loses.
  The assignments $\{u\mapsto 1\}$,$\{e\mapsto 1\}$ represent a game where the
  existential player wins. There exists a winning strategy for the existential
  player, which is $\{e\mapsto u\}$.
   Hence, the formula is true.
\end{example}

\begin{example}
  Consider the formula
  $\exists e_1\forall u\exists e_2.\,\{
  e_1\lor u\lor e_2,
  e_1\lor u\lor \lnot e_2,
  \lnot e_1\lor \lnot u\lor e_2,
  \lnot e_1\lor \lnot u\lor \lnot e_2\}$.
  The assignments $\{e_1\mapsto 0\}$, $\{u\mapsto 0\}$,$\{e_2\mapsto 1\}$ represent a game where the
  existential player loses.
  There exists a winning strategy for the universal player, which is $\{u\mapsto e_1\}$.
   Hence, the formula is false.
\end{example}

The game-theoretic perspective lets us look at the selection-based solving symmetrically:
the existential player tries to deselect clauses---so that the matrix
becomes true---and the universal tries to select clauses---so that the matrix becomes false.

The following concepts concretize this idea.
For each clause $C\in\phi$ and a
quantification level $k\in 1..n$ we introduce a variable~$s_C^k$, which indicates that the clause
$C$ is \emph{selected at the quantification level $k$}. If $s_C^k$ is false, then we say that
$C$ is \emph{deselected at the quantification level~$k$}. Selecting and deselecting clauses is
guided by the following rules:
\begin{enumerate}
  \item A clause can be selected  at level $ k $ if it was selected in all the previous levels and
    all its literals at the level $ k $ are set to false.

  \item A clause can be deselected at level $k$ if it was deselected in one of the previous
    levels \emph{or} at least one of the literals at level $k$ is set to true.
\end{enumerate}

Observe that deselection behaves monotonically, i.e., once a clause is deselected it cannot be
selected back again. This is unsurprising as once a clause is satisfied  during the game, it
remains satisfied.

The next thing to consider is how to encode the rules of the game.
These are formulated as follows:
\begin{enumerate}
\item If all clauses $C\in\phi$ are deselected at some level~$k$, the universal player loses.
\item If a clause $ C\in\phi $ does not contain any literal $ l $ with $\lev(l)>k$ then the existential
   player loses if the clause is selected at level~$ k $.
\end{enumerate}

These constraints correspond to the rules discussed above but they detect as soon as possible
that a player has lost, i.e.,
once all clauses are satisfied, the universal player loses,
and
once the empty clause is derived, the existential player loses.

In  its workings, the proposed algorithm is similar to the two-level one.
Starting from the first level, the players select and deselect clauses based on the rules of the
game.
At each level~$i$ we maintain a condition $\condition_i$, which blocks choices that
lead to a loss of player~$Q_i$. These conditions are strengthened throughout the course of the algorithm.
If a condition~$\condition_i$ becomes unsatisfiable,  it means that
player~$Q_i$ has \emph{lost} under
the current choices. Hence, we perform an analysis that tries to rectify some of the
previous choices of the losing player. This consists in
strengthening  one of the conditions~$\condition_k$ for some~$k<i$
(we look more closely at the loss analysis later).

\begin{algorithm}[t]
  \lFor{$i\in 1..n $}{$\condition_i\gets \InitCondition(i)$}
  $\round\gets 1$\;
  $S_0\gets\phi$\;
  \While{true}{\label{ln:qbfwb}
    \If{$\round=n+1$}{
      $\conflict\gets\comprehension{s^n_C\mapsto 0}{C\in\phi}$\;
      \Goto~\ref{line:analyze}\label{ln:qbfgoto}\;
    }
    $\alpha\gets\bigwedge_{C\in S_{\round-1}} s_C^{\round-1}
    \land\bigwedge_{C\notin S_{\round -1}} \lnot s_C^{\round-1}$\;
    $(\mu,\conflict)\gets\SAT(\condition_{\round}\land\alpha) $\label{ln:qbfmodel}\;
    \If{$\mu=\nul$} {
      $(\btround,C_L)\gets\analyze(\round,\conflict)$\label{line:analyze}%
      \label{ln:analyze}\;
      \lIf{$\btround = -1 $}{\Return $(Q_{\round}=\forall)$\label{ln:qbfterm}}
      $\condition_{\btround}\gets\condition_{\btround}\land C_L$\;
      $\round\gets\btround$%
      \label{ln:backtrack}\;
    }\Else {
      $S_{\round}=\comprehension{C\in\phi}{\mu(s_C^{\round})=1}$\;
      $\round\gets\round+1$\;
    }\label{ln:qbfwe}
  }
  \caption {\Qesto: Solving PQCNF}\label{algorithm:qbf}
\end{algorithm}
\begin{algorithm}[t]
\Func \InitCondition($k$)\;
  \lIf{$k=1$}{
    $\condition_1\gets\comprehension{s_C^1\triangleq\bigwedge_{l\in C,\lev(l)=1}\lnot l}{C\in\phi}$
  } \lElse{
    $\condition_k\gets\comprehension{s_C^k\triangleq s^{k-1}_C\land\bigwedge_{l\in C,\lev(l)=k}\lnot l}{C\in\phi}$
  }
  \lIf{$Q_k=\forall$}{$\condition_k\gets\condition_k\land\bigvee_{C\in\phi} s_C$}
  \Else{
    \For{$C\in\phi$}{
     \If{there is no existential $ l\in C $ s.t.\  $\lev(l)>k$}{
   $\condition_k\gets\condition_k\land s^k_C $}}
   }
   \Return $\condition_k$\;
   \caption {Condition Initialization}\label{algorithm:initialization}
\end{algorithm}
We name the algorithm \Qesto ({\it Qbf clausE SelecTion sOlver}), which
is listed in Algorithm~\ref{algorithm:qbf}.
The algorithm uses the subroutine \InitCondition (Algorithm~\ref{algorithm:initialization}) to initialize the conditions $\condition_i$,
$i\in 1..n$.
Then \Qesto continues by iterating the main loop
(lns.\,\ref{ln:qbfwb}--\ref{ln:qbfwe}) until the formula is solved.
In each iteration
the algorithm operates at a quantification level $\round$.  It is constructing sets $ S_i$,
which consist of the clauses currently selected at level~$i$.  To simplify the presentation, we
also introduce the set~$S_0$, which contains all the clauses in $\phi$. To construct the set
$S_{\round}$ we search for a model of the condition $\condition_{\round}$
restricted by the selections made so far (ln.\,\ref{ln:qbfmodel}). If this SAT call results in unsatisfiability, we
backtrack to a quantification level chosen by the function
$\analyze$. If the analysis fails, it returns the backtracking level~$-1$
and the algorithm terminates---the formula is true iff the losing player is the universal player
(ln.\,\ref{ln:qbfterm}).
  One special case is treated at the beginning of the loop and this is
  when the quantification level is $n+1$ (ln.\,\ref{ln:qbfgoto}). This means that the existential player satisfied
all the clauses (recall that $Q_n=\exists$), in which case we treat this situation as losing
for the universal player.

\subsection{Loss Analysis}
\begin{algorithm}[t]
  \caption{Loss analysis for the existential player\label{algorithm:exists}}
\Func \analyze(\round,\conflict)\;
\Begin {
   $S_c\gets\comprehension{C\in\phi}{\lnot s^{\round-1}_C\in\conflict}$\;
   $E\gets\comprehension{l}{l\in C, C\in S_c, l\text{ existential},\lev(l)<\round}$\;
   \lIf{$E=\emptyset$}{\Return$(-1,\cdot)$}
   $\btround\gets\max{\comprehension{\lev(l)}{l\in E}}$\tcp*[r]{btrack lev}
   $C_L\gets\bigvee_{C\in S_c}{\lnot s_C}$\tcp*[r]{learned clause}
   \Return $(\btround, C_L)$
}
\end{algorithm}

Let us look more closely at how \Qesto recovers from a
loss of either of the players (lns.\,\ref{ln:analyze}--\ref{ln:backtrack}).
Hence, we consider  the situation where one of the conditions $\condition_i$
becomes unsatisfiable under the choices of selected clauses made by the players so
far. This situation is analyzed and two outcomes are computed: the
quantification level~$\btround$ where the algorithm returns to, and, a
strengthening (a \emph{learned clause}) for the condition~$\condition_{\btround}$ that will prevent the
losing player from making the same mistake again. This is somewhat analogous
to how SAT solvers perform conflict analysis.

To get the best out of the underlying SAT solver, we rely on the
ability of the SAT solver to
provide us with a \emph{final conflict clause} for unsatisfiable calls.  For a formula
$\psi$ and a conjunction of literals
$\alpha$---designated as \emph{assumptions}---\emph{final conflict clause}
\conflict is a clause that contains only  variables appearing in the
assumptions~$\alpha$ and it holds that $\psi\models\conflict$.  Which practically means that if
$\psi\land\alpha$ is unsatisfiable, then $\psi\land\lnot\conflict$ is also unsatisfiable and
$\lnot\conflict$ is a subset of $\alpha$ that is sufficient to arrive to unsatisfiability.

In our scenario, at level $\round$, the assumptions are made up of
the variables $s_C^{\round-1}$ and these signal to the SAT solver what choices were made so
far.  Hence, inspecting the clause~$\conflict$  lets us infer with more precision which choices
were responsible for the loss of the player at hand.\footnote{%
  This approach also enables us to use \emph{incremental SAT}~\cite{een-bmc03}
so that  only the assumptions need to change when the same level is reached again.}

Note that  instead of the  final conflict clause technique, we could have
analyzed the corresponding \emph{resolution refutation}---if the solver produces one~\cite{zhang-date03}
(in fact only the leaves of the refutation are needed, similarly as in the \emph{core extraction}
techniques, which appear in a number of applications of SAT).

As the loss analysis is quite  different for the two players, we separate it into two respective
procedures.  The analysis for the existential player is quite straightforward and is presented in
Algorithm~\ref{algorithm:exists}. The final conflict clause~$\conflict$  gives us  a set of selected
clauses $S_c$ at level $\round-1$ that lead to the unsatisfiability of
$\condition_{\round}$. To rectify this, the existential player must ensure that at least one of
these clauses is \emph{deselected}, the next time we get to the level~$\round$.   For
the  existential player to deselect one of the clauses at a level $l<\round$
there must be an existential literal at that level. Otherwise, the existential player cannot
influence the state of the clause.
Hence, the backtrack level is calculated as the maximum over the levels of  existential
literals in the clauses~$S_c$ that are at  a lower level than~$\round$. If there
are no such literals, then the existential player cannot rectify the loss  and the formula is solved
(it is false). The learned clause is constructed so that it forces the deselection of at least one clause from the
problematic set~$S_c$.

\begin{algorithm}[t]
  \caption{Loss analysis for the  universal player\label{algorithm:universal}}
\Func \analyze(\round,\conflict)\;
\Begin {
   $S_c\gets\comprehension{C\in\phi}{s^{\round-1}_C\in\conflict}$\;
   $S'_c\gets\comprehension{C\in S_c}{C \text{ not satisfied by an exist.\ lit}}$\;
   \lIf{$S'_C =\emptyset$}{\Return $(-1,\cdot)$}
   $\btround\gets\max{\comprehension{k}{s_C^k=0,C\in S'_c,k\text{ universal}}}$
   $R\gets\comprehension{C\in S_c}{C \text{ sat.\ by an ex.\ lit at lev.}>\btround}$\;
   \Return $(\btround, \bigvee_{C\in(S_c\smallsetminus R)} s_C^{\btround}) $
}
\end{algorithm}

The analysis for the universal player is inverse
(Algorithm~\ref{algorithm:universal}).
At the time of a loss, there is some set of clauses~$S_c$ that are deselected and are causing the
universal player to lose at the level $\round$.
If a  clause is  deselected by the existential player, the
universal player cannot do anything about it.
So we shift our attention to clauses $S_l'\subseteq S_c$ that
are deselected by the universal player, i.e.,  satisfied by a universal literal.
If there are no such clauses, the universal player cannot rectify the loss and
therefore the formula is true.
Otherwise, the backtrack level is computed as the
maximum over the levels where the clauses in~$S'_c$ were deselected.
Now we could define the learned clause as $\bigvee_{C\in S_c}{s_C^{\btround}}$,
but we can do better if we realize that the existential player can always repeat any choices that
he has made \emph{after}~$\btround$. Hence,  if the existential player was able to deselect
a clause after~$\btround$, he can do so again next time and therefore we remove these clauses
from the learned clause.

\subsection{Comparison to Other Approaches}\label{sec:others}

In the context of a CNF formula~$\psi$, a \emph{maximal  satisfiable set} (MSS) is a
set of clauses~$\psi'\subseteq\psi$ that is satisfiable and adding any clause from $\psi$ to it
makes it unsatisfiable.  A complement~$\psi\setminus\psi'$ of such set is called a \emph{minimal
correction set (MCS)}.  Conversely, \emph{minimal unsatisfiable set (MUS)} is a set
$\psi'\subseteq\psi$ that is unsatisfiable and removing any of its  clauses makes it satisfiable. It is well
known that the set of all MUSes is the set of all minimal hitting sets of MCSes
(and also the other way around).
This duality of hitting sets is useful for calculating all MUSes
or MUSes with certain properties, which has been exploited in a number of
works~\cite{liffiton-jar08,pms-aaai13}. Moreover, algorithms based on iteratively analyzing implicitly
represented  hitting sets find a growing number of
applications~\cite{karp-cpm10,karp-soda11,bacchus-cp11,stern-aaai12,karp-or13,bacchus-sat13,bacchus-cp13,slaney-ecai14}.
The two-level approach we propose  explicitly links this paradigm to QBF solving.
Indeed, solving a two-level QBF ($\forall\exists$) can be phrased as a search for an MUS
comprising of the clauses containing only the existential part. The universal part, imposes
``rules'' on which clauses may appear in an unsatisfiable set. Indeed,
any clauses $C_1,C_2$ that contain universal literals
$y\in C_1$, $\lnot y\in C_2$ are never considered together.  Algorithm~\ref{algorithm:2qbf} then
iterates over satisfiable subsets of the matrix~$\phi$ and requires that a complement of the satisfiable set
is selected next time, which is consistent with the idea that MUSes are hitting sets of MCSes.
Note that our algorithm does not to go over \emph{maximally} satisfiable sets but it only requires
satisfiable sets. Similarly, we do not require a \emph{minimal} unsatisfiable set, but
simply any unsatisfiable set. In our implementation, however, we make sure that the decision polarities of the
selection variables in the SAT solver are set so that the SAT solver tends to find larger
satisfiable sets.

The two-level approach %
can also be related to the counterexample guided abstraction refinement algorithm
AReQS~\cite{janota-sat11}. Indeed, the \emph{selector variables} in our approach
(Algorithm~\ref{algorithm:2qbf}) can be seen as the \emph{Tseitin variables} introduced in AReQS.
We should stress, however, that the generalization of AReQS--RAReQS~\cite{janota-sat12}--is already completely different from our generalization
(Algorithm~\ref{algorithm:qbf}). In RAReQS the generalization is done by recursion and the
algorithm introduces a number of fresh variables throughout its course.
While \Qesto and RAReQS are similar for the two-quantification-level case,
they are vastly different in the general case.
Unlike RAReQS, \Qesto does not introduce fresh variables as the algorithm progresses---selector variables are introduced at the very beginning and their number is
bound. In this sense, our proposed algorithm  differentiates itself from other expansion-based
algorithms that convert the given formula into a SAT problem~\cite{Benedetti04,Biere04}.
Further, RAReQS may create an exponential number of solvers
due to the recursive calls while in \Qesto, the number of solvers is always linear in the number of
quantification levels (and it is fixed).

In its skeleton,  our algorithm is similar to the \emph{conflict/solution-driven learning}
algorithms~\cite{giunchiglia-ijcai01,Zhang-iccad02}; in the sense that it starts from the outermost quantification
level, progresses to the inwards quantification level and backtracks whenever needed.
It is, however, very different in how the algorithm is constructed and how learning is performed.
Indeed, the building blocks of \Qesto are a SAT solver and selection variables, which let us navigate the
search space. This approach lets us devise a learning procedure  anchored in the SAT solver's
ability to analyze unsatisfiability (the final conflict clause),  rather than building a
dedicated unit propagator and learner; this gives us a strong engineering advantage.
\section{Experimental Evaluation}\label{section:results}
\begin{figure}[t]
   \includegraphics{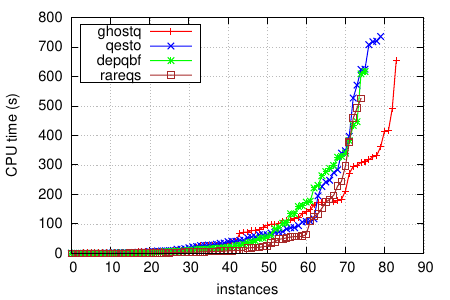}
   \caption{QBFLIB Benchmarks with  60\,s cutoff}\label{figure:eval12}
\end{figure}

\begin{figure}
  \includegraphics{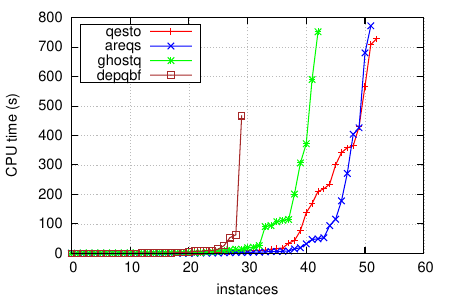}
  \caption{2QBF Benchmarks}\label{figure:2QBF}
\end{figure}

\begin{table}[t]
  \begin{tabular}{|l|c|c|c|c|}
    \hline
    & ghostq & qesto & depqbf & (r)areqs \\ \hline\hline
    \textit{QBFLIB} (276) & \textbf{137}    &  133  &  128   & 129    \\ \hline
    \textit{2QBF} (65)    & 43     &  \textbf{53}  & 30 &  52     \\ \hline
  \end{tabular}
  \caption{Number of Solved Instances}\label{table:solveds}
\end{table}

We have implemented a prototype of Algorithm~\ref{algorithm:qbf} named \textsf{Qesto}.  The
prototype used for the evaluation was implemented in $C^{++}$ and the SAT solver \textsf{minisat
2.2}~\cite{EenS03} was used.  The experimental results were carried out on Linux
machines with Intel Xeon 5160 3GHz processors and 4GB of memory. The time limit was set to
$800$\,s and the memory limit to~2GB.

For the evaluation we used the \emph{QBFLIB} benchmark suite used in the  \emph{2014 QBF
Gallery}\footnote{%
\url{http://www.kr.tuwien.ac.at/events/qbfgallery2013/benchmarks/eval2012r2.tar.7z}} and the
benchmarks from the \emph{2QBF track} of the \emph{2010 QBF Evaluation}.
All the instances were preprocessed by the preprocessor
\textsf{bloqqer}~\cite{biere-cade11} and instances solved by the preprocessor were excluded from
the evaluation.
For comparison we have
chosen a set of representative state-of-the-art QBF solvers.  The conflict/solution-driven solver
\textsf{DepQBF}~\cite{Lonsing-jsat10}; the CEGAR-based solver RAReQS and its 2QBF version
AReQS~\cite{janota-sat11,janota-sat12}; and the non-CNF solver GhostQ~\cite{klieber-sat10}.  Since
GhostQ is a non-CNF solver,  it was run on the unpreprocessed benchmarks as  the solver's reverse
engineering procedure benefits from their structure; it is also how it was run in the
competition.%

Table~\ref{table:solveds} summarizes the number of solved instances for the  considered solvers.
Figures~\ref{figure:2QBF} and~\ref{figure:eval12} show cactus plots for the respective benchmark
sets (cactus plots consist of points~$(x,y)$ meaning there are $x$ instances solved in $y$\,sec).
Since the QBFLIB cactus plot has a rather a long tail, the plot contains only those instances that
were \emph{not} solved within 60 seconds by all the solvers.

GhostQ clearly dominates in the QBFLIB  suite and it is followed by our prototype Qesto.
There is  a difference of only one solved instance between the solvers DepQBF and RAReQS.
The success of the non-CNF solver GhostQ, confirms the well-known fact that
representing QBF problems as CNF is harmful for solving~\cite{ansotegui-aaai05,klieber-sat10,goultiaeva-aaai10}
(we return to the topic in our future work discussion).

In the 2QBF suite  our prototype Qesto solves the most instances.
However, AReQS solves just~1 instance less, which comes as no surprise
as the algorithms are closely related
(see~Section~\ref{sec:others}).  Interestingly, GhostQ performs rather poorly on this set of
benchmarks, which suggests that our prototype gives a higher degree of
robustness when compared to others.
More detailed presentation of the results can be found on the authors' website\footnote{%
{\sf http://sat.inesc-id.pt/\%7Emikolas/sw/qesto/}}.

\section{Summary and Future Work}\label{sec:conclusions}

This paper describes an algorithm \Qesto to solve QBF in prenex form with a matrix in CNF.  The
algorithm is first presented in its two-level form ($\forall\exists$), which is shown to have a
clear correspondence to implicit hitting sets, which appear in a number of
applications of artificial intelligence. This two-level approach is also linked to the recent
CEGAR-based two-level QBF algorithm AReQS~\cite{janota-sat11}. The second half of the  paper
generalizes the two-level approach to QBFs with  arbitrary number of quantification levels,
giving rise to the \Qesto algorithm.
\Qesto is an entirely novel approach to QBF solving and the experimental evaluation suggests that it
is quite robust.  It should also be noted that \Qesto is quite  advantageous from the
engineering perspective since a SAT solver is utilized in a blackbox fashion.  \Qesto is
also quite interesting from the  theoretical perspective as it can be seen as a generalization of
duality hitting set principle for PSPACE.

On the QBFLIB benchmarks \Qesto was outperformed by the non-CNF solver
GhostQ~\cite{klieber-sat10}. This suggests that it would be worthwhile investigating how \Qesto
could be applied to non-CNF formulas. We believe that this should be possible as this was already
successfully done for conflicts/solution-driven learning algorithms~\cite{zhang-aaai06,goultiaeva-date13}.
Other improvements, such as \emph{pure literals} or \emph{variable dependencies}, are also considered for future
work.
\bibliographystyle{named}

\end{document}